# An Effective Hamiltonian for Symmetric Diarylmethanes from a Series of Analogous Quantum Chemical Models

*Seth Olsen and Ross H. McKenzie*

School of Mathematics and Physics, The University of Queensland, Brisbane QLD 4072 Australia

seth.olsen@uq.edu.au

**Abstract** We propose a single effective Hamiltonian to describe the low-energy electronic structure of a series of symmetric cationic diarylmethanes, which are all bridge-substituted derivatives of Michler's Hydrol Blue. Three-state diabatic Hamiltonians for the dyes are calculated using four-electron three-orbital state-averaged complete active space self-consistent field and multi-state multi-reference perturbation theory models. The approach takes advantage of an isolobal analogy that can be established between the orbitals spanning the active spaces of the different substituted dyes. The solutions of the chemical problem are expressed in a diabatic Hilbert space that is analogous to classical resonance models. The effective Hamiltonians for all dyes can be fit to a single functional form that depends on the mixing angle between a bridge-charged diabatic state and a superposition representing the canonical resonance. We find that the structure of the bridge-charged state changes in a regular fashion across the series. The change is consistent with an inversion of the sign of the charge carrier on the bridge, which changes from an electron pair to a hole as the series is traversed.

## 1. Introduction

Tri- and di-arylmethane dyes related to Michler's Hydrol Blue (c.f. Scheme 1) are some of the oldest molecules to be synthesized by humankind, and were key players in the development of industrial chemistry.[1, 2] Recent interest in these dyes has focused on the environmental sensitivity of their optical response.[3-11] In particular, a common binding-dependent fluorescence enhancement makes these dyes useful biological markers.[4-6, 12] The fluorescence enhancement is attributed to suppression of a competing double-bond isomerization reaction, the rate of which is influenced by the local viscosity.[13-16]

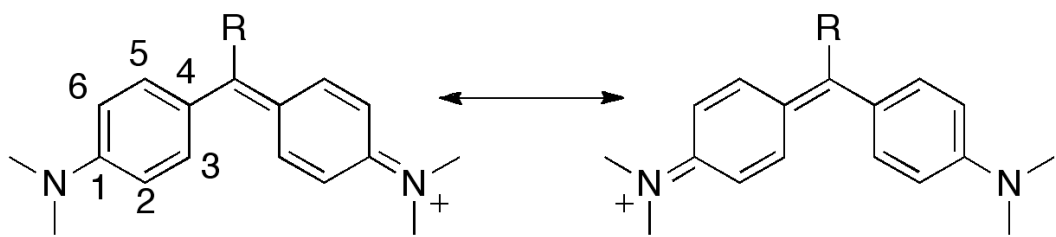

R= H (Hydrol Blue; HB)
   $CH_3$ ($\alpha$-methyl HB; MHB)
   $NH_2$ (Auramine-O; AO)
   OH (Michler's Ketone Conj. Acid; MKA)
   Ph (Malachite Green; MG)
   CN ($\alpha$-cyano HB; CNHB)
   F ($\alpha$-fluoro HB; FHB)

**Scheme 1.**

As expected from their long history of use, the low-energy electronic excitations of dyes such as in Scheme 1 have been subject to numerous modeling studies. A consistent theme is the notion of resonance between "canonical" valence-bond states with opposing charge localization and bond alternation (c.f. Scheme 1).[17-24] In some models, the Hilbert space is extended to cover "intermediate" structures, with charge on the bridge, which mediate superexchange between the canonical structures.[17, 18, 23, 24] The inclusion of intermediate structures is motivated by the expectation that the one-particle inter-ring transfer matrix element should scale as the overlap between ring frontier orbitals, and will be too small to account for the observed strong absorption in the visible spectrum.[17, 18] Alternatively, it has been noted that the inclusion of these states facilitates the quantitative parameterization of the dipole observables associated with the low-lying transitions.[23, 24] In any case, the charge carrier on the bridge in the "intermediate" structures is frequently taken to have the character of a hole for cationic dyes such as in Scheme 1.[17, 18, 23]

In this paper, we will present and analyze a series of diabatic[25] effective Hamiltonians for the low-energy spectrum of dyes in Scheme 1. The Hamiltonians are extracted from state-averaged complete active space self-consistent field (SA-CASSCF) and multi-state, multi-reference perturbation theory (MSRS2) calculations. A key element is the identification and exploitation of an isolobal analogy[26] between the active spaces of the SA-CASSCF solutions for the different dyes in Scheme 1. This allows the representation of the many-electron states in analogous diabatic states with a valence-bond[27] structure. The structure of the diabatic representations is identified with the structure of the Hilbert space of classical resonance-theoretic[17, 18, 28] and modern essential-state[23, 24] models. We will show that the charge carrier on the bridge in the intermediate state changes its sign as the molecular series is traversed. The evolution of the sign of the charge carrier can be described by a one-parameter family of model Hamiltonians obtained from the electronic structure computations.

## 2. Methods

### *2.1. Computations*

For each dye in Scheme 1, we obtained a ground state minimum geometry using an MP2[29]

model with a cc-pvdz basis set[30] (MP2//cc-pvdz). The Cartesian coordinates of these geometries are available in an online supplement. The geometries of all dyes were non-planar. This contradicts some assertions that have been made in the previous literature concerning Michler's Hydrol Blue.[31, 32] The total twist of the rings across the methine bridge was $30^0$-$40^0$. Where the bridge substituent was compatible with $C_2$ symmetry of the Hydrol Blue skeleton, the ground state geometry had this symmetry. Even in the symmetric cases there are two rotamers corresponding to different handedness of the twist of the bridge; these forms should exchange slowly due to the high barrier imposed by steric interaction of the rings. We worked with the right-handed rotamers.

For each dye at its ground state minimum, we obtained a solution of the four electron, three orbital, three state-averaged complete active space self consistent field[33] problem, using the same basis set (SA3-CAS(4,3)//cc-pvdz). Equal weights were applied to the three states in the average. The SA3-CAS(4,3) solutions are for the dye family are analogous (see below). The initial orbitals for the SA-CASSCF procedure were the UHF charge-natural orbitals[34] obtained for the doubly cationic (i.e. ionized) radical species.

We used the SA-CAS(4,3) adiabatic states for each dye as a reference basis for a multi-state multi-reference second-order Raleigh-Schrödinger perturbation theory (MSRS2) model. The zeroth-order Hamiltonian was defined on the Boys localized orbitals in the active space, and the state-averaged canonical orbitals outside of the active space. We verified that the results did not depend on whether the perturbation was applied in the Boys or state-averaged canonical representation of the active space. Only 32 active and closed orbitals were correlated in the perturbation theory. Since there are more orbitals uncorrelated for larger dyes, the error introduced by this is size-dependent. Based on previous calculations on an analogous 2-state averaged model (SA2-CAS(4,3)), Olsen determined that including all orbitals in the valence lowered the first excitation energy by ~0.1-0.15eV for all dyes except the largest (Malachite Green), for which the error was larger (~0.2eV).[35]

This is essentially the same protocol that we have used[36, 37] in our previous studies of the chemically related GFP chromophore system. The adiabatic excitation energies and transition-dipole observable of this model have been discussed in a previous paper. There, it was concluded that the SA3-CAS(4,3) model constitutes a balanced low-rank description of the low-lying excitations associated with the charge resonance.[38] The results were found to be insensitive to basis set size, and enlargement of the active space. Expansion of the active space was found to degrade the description of the charge difference densities associated with the excitation, for any level of description between the four-electron, three-orbital active space and the full $\pi$-electron active space.

The MSRS2 and SA-CASSCF adiabatic states were transformed into a quasi-diabatic representation using the block-diagonalization technique described by Pacher and Cederbaum.[39] The target space of the block-diagonalization was spanned by the "covalent" configurations over the Boys-localized active orbitals (c.f. Figure 1, bottom left). These configurations have one doubly filled orbital and one singlet pair distributed between the remaining two. This procedure produces diabatic states that are the same when either the SA-CASSCF or MSRS2 eigenvectors are used, because the latter two are related by a unitary transformation. It follows that the block-diagonalization procedure produces the same set of quasi-diabatic states in both cases. However, the transformation mapping the diabatic states onto the adiabatic states will be different in the SA-CASSCF and MSRS2 cases.

The state-averaged ensemble density matrix is invariant to rotations within the state-averaged space when the weighting is uniform over the space (as it is here).[40, 41] It follows that the state averaged ensemble density matrix is the same in the adiabatic and diabatic state bases (it is 1/3 times the identity in either representation).

The block-diagonalization algorthithm has been used in other work to produce diabatic states for single molecules, in order to analyze nonadiabatic reaction mechanisms.[39, 42, 43] We are not aware of other work where it has been used in conjunction with an isolobal analogy to directly analyze the electronic structure of families of similarly substituted molecules (although this does seem to us to be an obvious application of the technique).

*2.2. Analytical Framework*

In this section, we describe a general analytical diabatic-state model Hamiltonian for a symmetric dye of the type indicated in Scheme 1. This model will provide a framework for the interpretation of the quantum chemical results, so it is appropriate to introduce it now. The model is similar in spirit to classical resonance-theoretic models of cyanine-like (methine) dyes discussed by Moffitt[18], Simpson[28], Brooker[19] and Platt[20], and to recent essential-state models of polymethine[23], donor-π-acceptor quadrupolar[24], and triarylmethane[44] dyes that have been developed and applied by Painelli and coworkers. Our contribution here is to show that the structure of the Hilbert space invoked in these models (originally based on chemical intuition) can be justified as an approximation to the solution of a well-defined quantum chemical problem.

The electronic Hamiltonian is defined on a basis of diabatic states, labeled $|B\rangle$, $|L\rangle$ and $|R\rangle$. We parameterize the electronic Hamiltonian as

$$\langle L|\hat{H}|L\rangle = \langle R|\hat{H}|R\rangle = 0 \tag{1a}$$

$$\langle B|\hat{H}|B\rangle = \varepsilon \tag{1b}$$

$$\langle B|\hat{H}|L\rangle = \langle B|\hat{H}|R\rangle = -\tau \tag{1c}$$

$$\langle L|\hat{H}|R\rangle = -\beta \tag{1d}$$

Here the energies of the (degenerate) states $|L\rangle$ and $|R\rangle$ have been set to zero, $\varepsilon$ is the energy of $|B\rangle$ relative to these, $\tau$ is the coupling between the extreme states $|L\rangle$ and $|R\rangle$ to $|B\rangle$, and $\beta$ is the direct coupling between the states $|L\rangle$ and $|R\rangle$. This is the most general three-state Hamiltonian that is consistent with the $C_2$ symmetry of the common bis-$N,N'$-dimethylaniline skeleton.

The Hamiltonian (1) can be diagonalized in two steps. The first step involves a rotation by $\pi/2$ in the subspace $(|L\rangle,|R\rangle)$. The mixing angle in this subspace is dictated by symmetry, and produces superposition states $|L \pm R\rangle = (|L\rangle \pm |R\rangle)/\sqrt{2}$ adapted to the $C_2$ molecular symmetry. The second step

involves mixing the intermediate state $|B\rangle$ with the *even* combination of extreme states (i.e. $|L + R\rangle$) to produce the final adiabatic states. This second transformation is characterized by a mixing angle $\theta$.

$$\theta = \frac{1}{2}\tan^{-1}\left(\frac{-2\sqrt{2}\tau}{\varepsilon+\beta}\right) \qquad (2)$$

This mixing angle is a characteristic of each dye. Since the only chemical difference between them is the bridge subsituent, we can consider it as a characteristic of the substituent. We will find it useful to use it as an independent variable in our analysis of the electronic structure of the dye series.

### 3. Results

#### 3.1. Isolobal and Isoconfigurational Analogies

There is an isolobal (orbital) analogy[26, 45] amongst the active orbitals of the SA3-CAS(4,3) solutions for the dyes in Scheme 1. This analogy is shown in Figure 1. In the Boys-localized active space, there are orbitals localized on each of the two rings and the bridge. A similar analogy can be found for the two-state averaged SA2-CAS(4,3) problem for these same dyes.[35]

The isolobal analogy establishes a mapping between the active spaces of the dyes in Scheme 1. This can also be leveraged to generate an iso*configurational* analogy mapping the configurational spaces of the different dyes onto one another. This analogy is highlighted in the bottom left corner of Figure 1.

The isolobal and isoconfigurational analogies are chemically meaningful only so far as the orbitals localize in a similar sense for each dye. The Boys localization procedure was used. This localization algorithm uses a physically well-defined objective function corresponding to the distance between the charge centres of the different orbitals.[46] The observed fact that it produces a similar localization in every case is a reflection of the chemical similarity between the dyes in the set. The localization performs reasonably well for all dyes studied, but is not perfect in any case. Some of this will be due to the requirement of orthogonality. We note that the corresponding orbitals of the two-state problem show a cleaner localization. The orbitals for Malachite Green (X=–$C_6H_5$) shown in Figure 1 are less localized than in other cases, as judged by the eye. This indicates that Malachite Green is a more rigorous test of our analysis than the other dyes here.

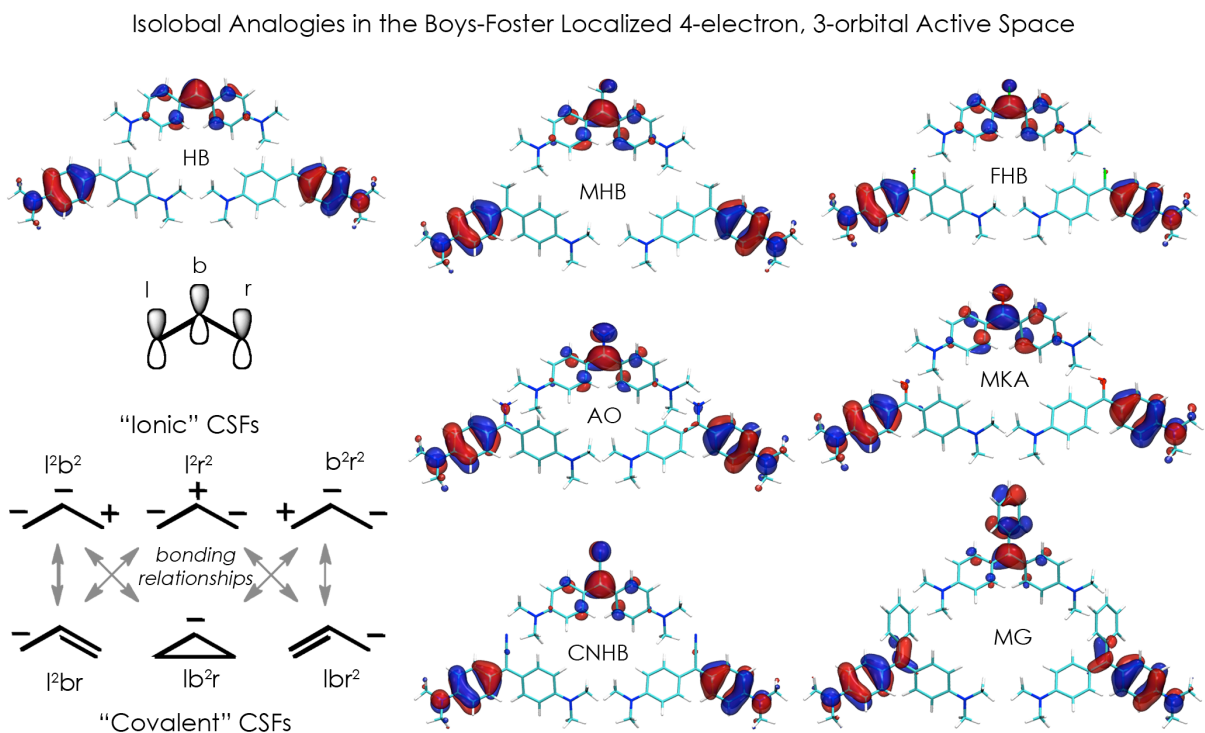

**Figure 1.** Boys-Localized active orbitals of bridge-substituted derivatives of Michler's Hydrol Blue. The localized orbitals are analogous to the p orbitals of an allylic anion (middle left). The orbital analogy induces an analogy between singlet configurations constructed over the orbitals (lower left).

*3.2. Diabatic States*

Exploiting the isolobal and isoconfigurational analogies described above, we can define an analogous set of quasi-diabatic (from here on, just "diabatic") states by block-diagonalization of the Hamiltonian[39] in the basis of analogous singlet configurations generated by the localized active orbitals. The structure of the diabatic states in the basis of configurations is shown in Figure 2. The target space for the block-diagonalization was the 3-dimensional subspace spanned by the "covalent" configurations, as indicated in the bottom left of Figure 1. The diabatic states are the same for the MSRS2 and SA-CASSCF calculations, because they depend only on the projection of the space spanned by the adiabatic states onto the configuration basis[39]; they are invariant to the unitary transformation induced by the multi-state MSRS2 treatment.[47]

There is a clear identification between the diabatic states in Figure 2 and the resonating bond structures that have served as the conceptual basis for several models of organic molecular optoelectronic properties.[17-24, 28] In particular, the diabatic states labeled $|L\rangle$ and $|R\rangle$, which are shown in the left and right panels of Figure 2, are analogues of the resonating structures in Scheme 1. To see this, note that they are dominated by the configurations $|l^2br\rangle$ and $|lbr^2\rangle$, and note also that they also carry significant contribution from the "ionic" configurations that represent stabilization of the singlets in the dominant configurations.[48] For $|L\rangle$, these are $|l^2b^2\rangle$ and $|l^2r^2\rangle$; for $|R\rangle$, these are $|b^2r^2\rangle$ and $|l^2r^2\rangle$. The contribution of these ionic structures indicates that the diabatic states $|L\rangle$ and $|R\rangle$ represent "normal" valence-bond structures, which should be amenable to established parametric descriptions.[27]

Note that the structure of the states $|L\rangle$ and $|R\rangle$ do not change much from dye to dye. In particular, the ordering of the contributions from the configurations varies smoothly, and the magnitude of the variation is modest on the scale of Figure 2.

If the states $|L\rangle$ and $|R\rangle$ are to be identified with the canonical resonating valence-bond structures (c.f. Scheme 1), then it is reasonable to suppose that $|B\rangle$ should be identified with the "intermediate" structure that has been invoked in resonance color theories.[17-20] In this case, Figure 2 highlights a significant point of departure from these models, because in this case the state is *not* the same for all of the dyes in the series. More specifically, Figure 2 shows a clear transformation over the series from a state dominated by the configuration $|lb^2r\rangle$ to one dominated by $|l^2r^2\rangle$. In other words, the sign of the charge carrier on the bridge in this diabatic state inverts its sign as the series is traversed.

The state $|B\rangle$ also changes in other ways. For example, there is a clear amplification of the amplitude of $|B\rangle$ on the "dipolar" covalent ($|l^2br\rangle$, $|lbr^2\rangle$) and ionic ($|l^2b^2\rangle$,$|b^2r^2\rangle$) configurations, but this is minor in comparison with the change in the dominant configuration. We also note that Malachite Green also appears exceptional, because the relative contribution of the dipolar covalent and ionic configurations is

inverted with respect to the broader trend.

The change in the character of the dominant configuration of $|B\rangle$ is the single greatest change that can be seen in Figure 2.

Note also that the ionic configurations that would indicate stabilization of the singlet pair in the configuration $|lb^2r\rangle$ have only small amplitude in $|B\rangle$. This implies that this component of $|B\rangle$ represents a biradical pair, not a stable chemical bond.

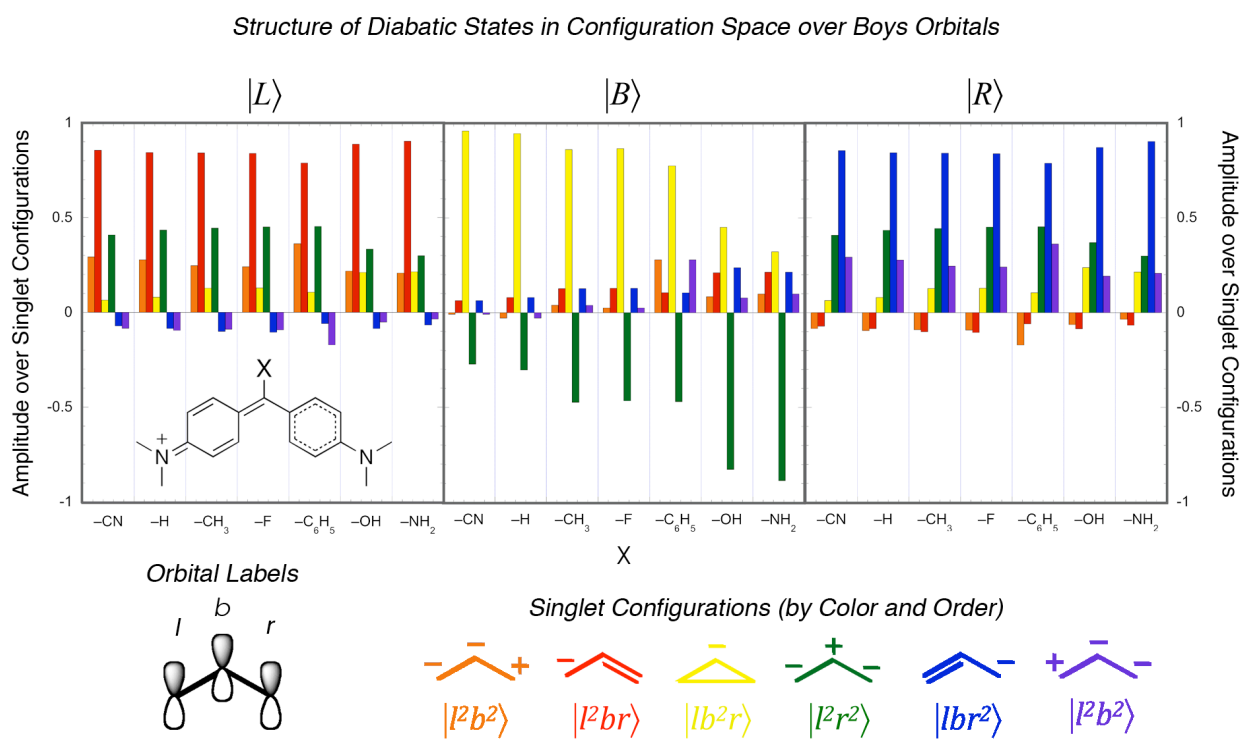

**Figure 2.** Amplitudes of block-localized diabatic states in the space of configurations over Boys-localized orbitals. Boys orbitals are shown in Figure 1. Amplitude bars are color-coded by configuration; color legend is shown at bottom. The left to right order for each dye is the same as that in the legend. Labels for orbitals and configurations are the same as indicated in Figure 1.

### 3.3. Diabatic Hamiltonians

The diabatic Hamiltonians calculated using the MSRS2 and SA-CASSCF models are shown in Figures 3 and 4, respectively, using the parameters $\varepsilon$, $\beta$ and $\tau$ defined in Equation (1) above.

Hamiltonians for asymmetric dyes (i.e. X=–CH$_3$, –OH) were symmetrized by averaging over the elements connected with the $|L\rangle$ and $|R\rangle$ diabatic states. The magnitude of the asymmetry for these elements is shown in Figures 3 and 4 using error bars. It is only visible to the eye for the case of X=–OH; even there it is small, so that the use of our symmetric analytical model is a reasonable approach even for the asymmetric dyes. The parameters are plotted against the mixing angle $\theta$ (c.f. Equation (2)), which is calculated from the parameters themselves.

The most important point made in Figures 3 and 4 is that the parameters $\varepsilon$ and $\tau$ change significantly over the set, while $\beta$ varies over a much smaller range.

The changes in $\varepsilon$ and $\tau$ are consistent with a rotation of the diabatic state $|B\rangle$ within a two-dimensional subspace characterized by the same rotation angle $\theta$ (c.f. Equation 2). To see this, presume that the state $|B\rangle$ can be expanded as

$$|B\rangle = \cos(\theta)|B-\rangle + \sin(\theta)|B+\rangle \quad (3)$$

We have used the same mixing angle $\theta$ that characterizes the mixing of $|B\rangle$ with the superposition $|L + R\rangle$. Given this, the on- and off-diagonal Hamiltonian elements connected with $|B\rangle$ can be written

$$\langle B|H|B\rangle = \cos^2(\theta)\langle B-|H|B-\rangle + \sin^2(\theta)\langle B+|H|B+\rangle \quad (4)$$

$$\langle B|H|X \in \{L,R\}\rangle = \cos(\theta)\langle B-|H|X\rangle + \sin(\theta)\langle B+|H|X\rangle \quad (5)$$

When these expressions are inserted into Equation (1), we obtain the functional form that was used to fit the parameters $\varepsilon$ and $\tau$ in Figures 3 and 4. As can readily be seen, these fitted forms describe the variation in parameters well (more specifically, the greatest absolute residual is 0.17eV, and is smaller than 0.1eV for almost every other case). This shows that the greatest variation seen in the parameters of the diabatic Hamiltonians can be accounted for by a rotation of $|B\rangle$ within a two-dimensional space, with the rotation characterized by the same mixing angle that characterizes the mixing of $|B\rangle$ with $|L + R\rangle$.

There are clearly other differences between the dyes that appear in Figures 3 and 4. We have not explained the variation in the parameter $\beta$, nor will we. The standard deviation in $\beta$ is 0.06eV for the MSRS2 case, and 0.19eV for SA-CASSCF. These energies are on or within the scale of error that is usually associated with multireference perturbation theory estimates of electronic excitation energies.[49] A detailed microscopic account of the variation of this parameter cannot be convincingly made using the data set we have. The approximation that $\langle L|H|R\rangle$ is independent of the nature of the bridge substituent is consistent with the approximate localization and transferability of the orbitals in Figure 1.

Using the functional fits in Figures 3 and 4, we can extract estimates of the limits of the parameters $\varepsilon$ and $\tau$. According to Equations (3) and (4), these can be associated with the energies of the hypothetical states $|B-\rangle$ amd $|B+\rangle$, and with their couplings to $|L\rangle$ and $|R\rangle$. The fits to the MSRS2 Hamiltonians indicate that $\langle B\text{-}|H|B\text{-}\rangle$ = 1.92 eV and $\langle B+|H|B+\rangle$ = -2.54 eV, while the fits to the SA-CASSCF Hamiltonians indicate 3.79 eV and -9.32 eV, respectively. This suggests that the MSRS2 perturbation changes the splitting of these states by a very significant amount. The fits also suggest that the coupling of $|B+\rangle$ with $|L\rangle$ and $|R\rangle$ is strong (2.13 eV for MSRS2, 3.32 eV for SA-CASSCF), while the coupling associated with $|B-\rangle$ is two orders of magnitude smaller (0.02 eV and 0.04 eV, respectively).

Taking into account the results shown in Figure 2, it is reasonable to think that the subspace spanned by $|B-\rangle$ and $|B+\rangle$ should have a significant contribution from the configurations $|lb^2r\rangle$ and $|l^2r^2\rangle$. Since the dyes for which $\theta\sim 0$ have $|B\rangle$ dominated by comparable contributions of these configurations with opposite phase, it would seem a good guess to make the association $|B–\rangle \sim |lb^2r\rangle – |l^2r^2\rangle$ and likewise $|B+\rangle \sim |lb^2r\rangle + |l^2r^2\rangle$. This would be consistent also with the fact that $|lb^2r\rangle$ and $|l^2r^2\rangle$ are coupled through a two-electron excitation, so that the perturbation treatment would shift the energy of the excitation significantly, as observed. It would also be consistent with the low magnitude of the coupling elements $\langle B–|H|L,R\rangle$ inferred from the fits in Figure 3, because the one-electron contributions to the coupling between dominant configurations will cancel (i.e. $\langle lb^2r|H|l^2br\rangle = \langle l^2r^2|H|l^2br\rangle \sim \langle b|f^{SA}|r\rangle$ where $f^{SA}$ is an effective one-particle Hamiltonian – here the state-averaged Fock matrix). This will be true for any approximate model where the correlation contributions are diagonal in the Fock space, which includes well-known semi-empirical Hamiltonians.[50]

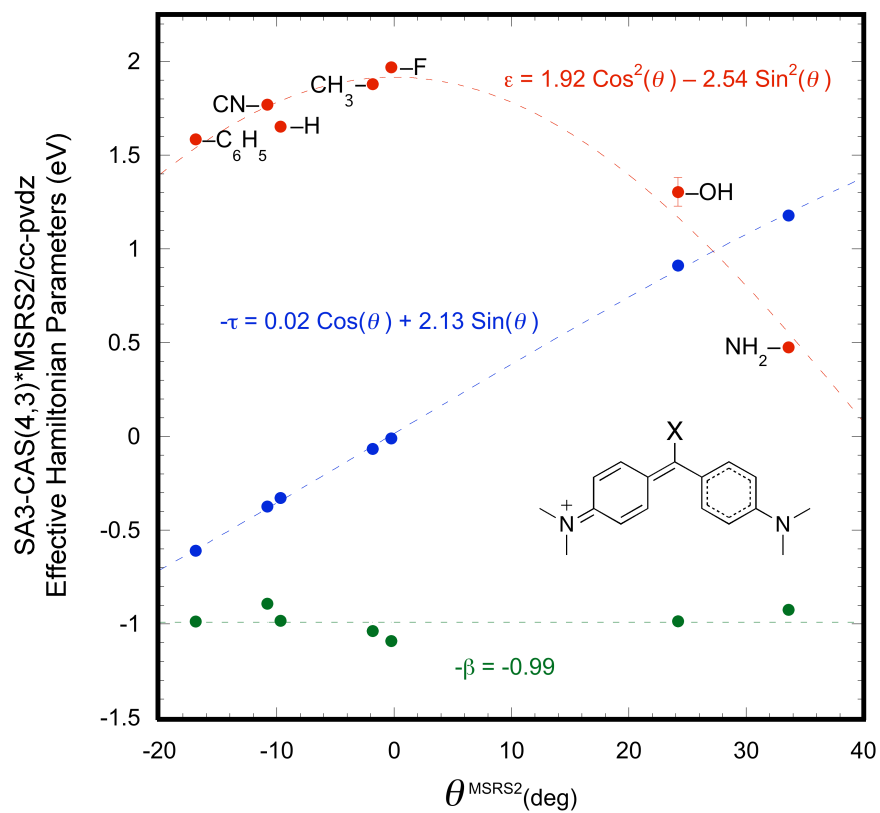

**Figure 3.** Effective Hamiltonian parameters for dyes in the set, extracted from SA3-CAS(4,3)*MSRS2/cc-pvdz quantum chemistry calculations. The parameters are plotted against the mixing angle $\theta$ (c.f. equation 2).

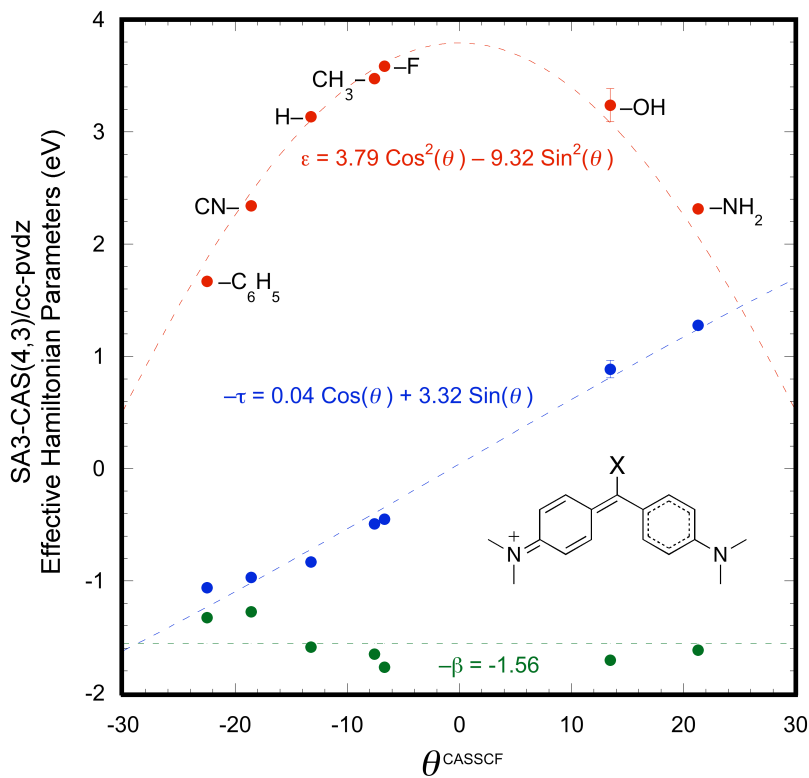

**Figure 4.** Effective Hamiltonian parameters for dyes in the set, extracted from SA3-CAS(4,3)/cc-pvdz quantum chemistry calculations. The parameters are plotted against the mixing angle $\theta$ (c.f. Equation 2).

*3.4. Adiabatic States*

The structure of the adiabatic states in the diabatic state basis, calculated using MSRS2 and SA-CASSCF is shown in Figure 5.

The structure of the states is broadly consistent with three-state resonance-theoretic models of color in cyanine-type chromophores.[18, 23] In particular, the ground state $|S_0\rangle$ has a large contribution from the even superposition $|L+R\rangle$ and a smaller contribution from $|B\rangle$, while the first excited state $|S_1\rangle$ is dominated by the odd superposition $|L-R\rangle$ and has insignificant contribution from $|B\rangle$ (for symmetric dyes, vanishing contribution). The second excited state $|S_2\rangle$ is dominated by $|B\rangle$, with only a very small contribution from states $|L\rangle$ and $|R\rangle$.

The sign of the mixing of $|B\rangle$ with $|L+R\rangle$ is positive for dyes with $\theta > 0$ and negative for $\theta < 0$ (this is implicit in the definition of $\theta$). The contribution of $|B\rangle$ to $|S_0\rangle$ is reduced by the application of MSRS2 for those dyes which have $\theta > 0$, and amplified for dyes with $\theta < 0$. With this in mind, a revisit to Figure 2 indicates that a major effect of the perturbation in MSRS2 is to lower the amplitude of $|lb^2r\rangle$ in the ground state of all dyes. The change in $\theta$ upon application of the MSRS2 treatment is shown in Figure 6, and is always positive.

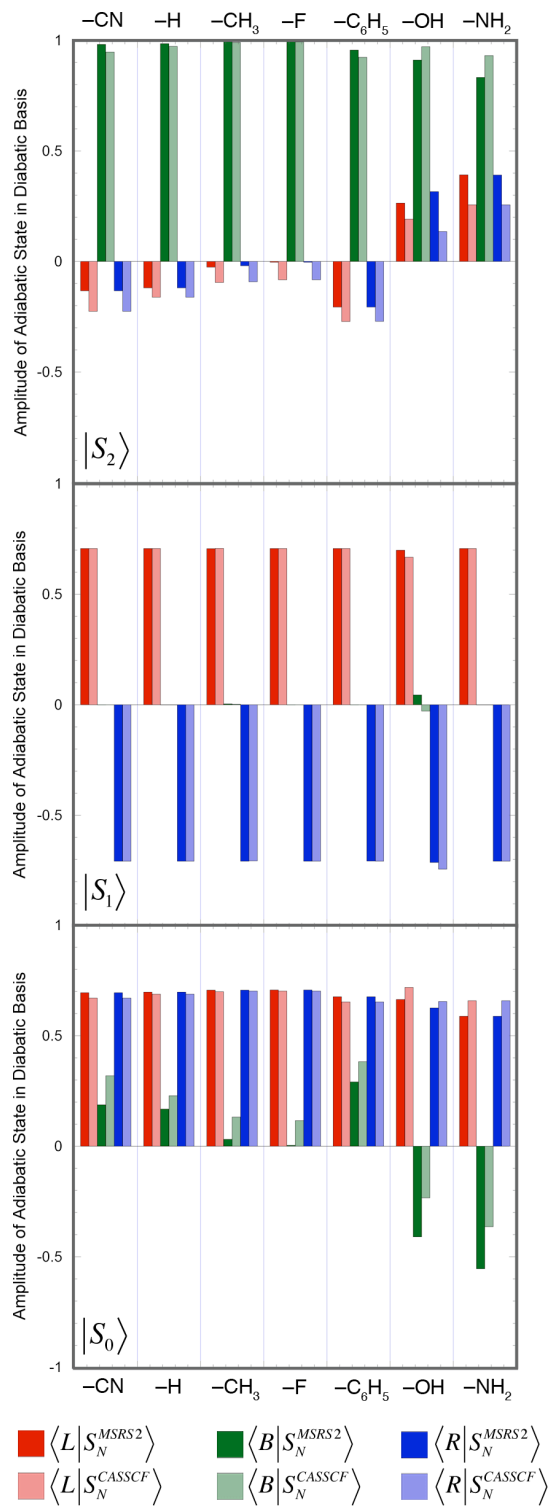

**Figure 5.** Adiabatic states represented on the diabatic basis. Adiabatic states calculated with MSRS2 (deep color) and SA-CASSCF (light color) are expressed through their amplitudes on the diabatic states $|L\rangle$ (red), $|B\rangle$ (green) and $|R\rangle$ (blue). Diabatic states are as shown in Figure 2, and are the same for both MSRS2 and SA-CASSCF calculations.

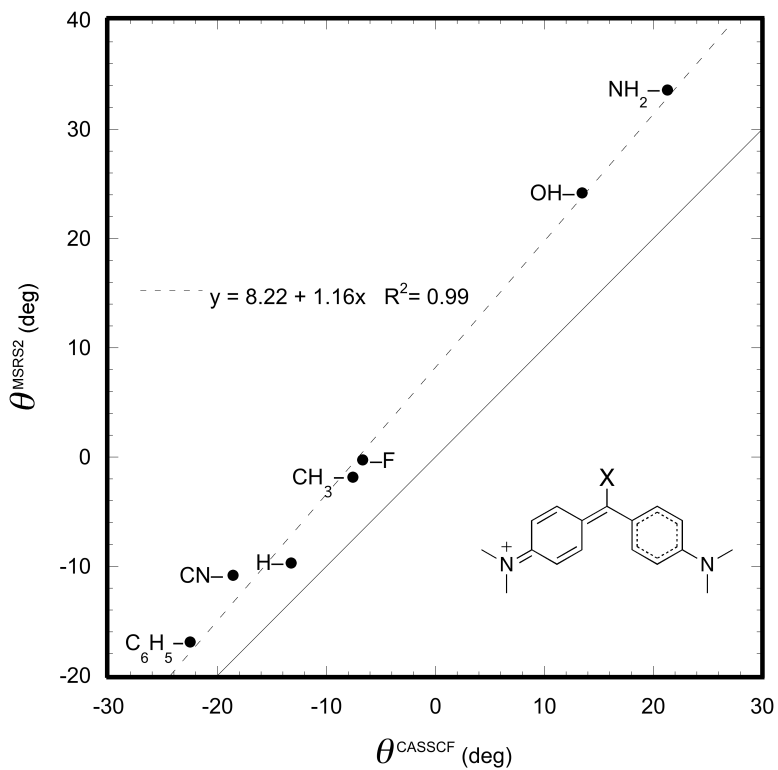

**Figure 6.** Comparison of the mixing angle $\theta$ (c.f. Equation 2) calculated using the SA-CASSCF (horizontal axis) or MSRS2 (vertical axis) adiabatic states. The dotted line represents identity between the two sets.

**4. Discussion**

We have shown that the solutions to a four-electron, three-orbital state-averaged complete active space self-consistent field and multi-state, multireference perturbation theory model can be cast in a diabatic form with a valence-bond structure identifiable with classical resonance models[17, 18, 20, 28] and modern essential-state models[23, 24, 44]. These models are built on a three-state Hilbert space spanned by two "canonical" resonating structures, which are conjugate in the sense of flipped polarity and bond alternation (c.f. Scheme 1), and by a third "intermediate" structure with charge on the bridge.

Despite the similarity in gross structure between our results and earlier resonance models, we have highlighted an important difference. Whereas earlier models usually assumed that the charge on the bridge is carried by a hole, our results indicate that the charge can be carried by an electron pair for certain members of the dye class. The charge carrier changes sign as the series is traversed. The progress of the sign change can be expressed as a rotation of the intermediate state in a two-dimensional subspace, with the rotation angle slaved to the angle $\theta$ that determines the mixing between the intermediate and the canonical resonant state. In many previous models, the assumption that the charge in the intermediate state had the same character as in the canonical states was consistent with the use of a single-particle model. For example, even though Pauling *reasoned* with valence-bond (i.e. many-body) structures akin to those in Scheme 1, the model he ended up proposing was essentially an orthogonal tight-binding model for the motion of a positive charge.[17] In Moffitt's treatment of the formamidinium ion, he implicitly used a complete active space expansion as we have done here, but then assumed that the effect of the structure with negative charge on the bridge could be treated as a perturbation.[18] Our results contradict this assumption.

The angle $\theta$ is broadly, but imperfectly, correlated with the Brown-Okamoto substituent parameter $\sigma_P^+$. This parameter is a measure of the ability of the subsituent to stabilize positive charge through conjugative interactions.[51] The correlation is shown in Figure 7. A visual inspection of the fits indicates that the poor quality of the regression is due to deviation in the specific cases of X = –CN and –$C_6H_5$. For X = –CN, the Brown-Okamoto $\sigma_P^+$ parameter[51] is the same as the Hammett parameter, and so it is arguable whether it is appropriate description of the channel for conjugative substituent effects relevant here. For X = –$C_6H_5$, the origin of the deviation is not so clear. Given that Malachite Green is also an outlier in terms of the orbital localization and the structure of the diabatic states, it may well be that the theoretical model we are using does not apply so well for this case. It seems that the best we can say is that the substituent channel is similar to the channel for electrophilic aromatic substitution, but only for cases where the substituent is not aromatic and where $\sigma_P^+ \leq 0$.

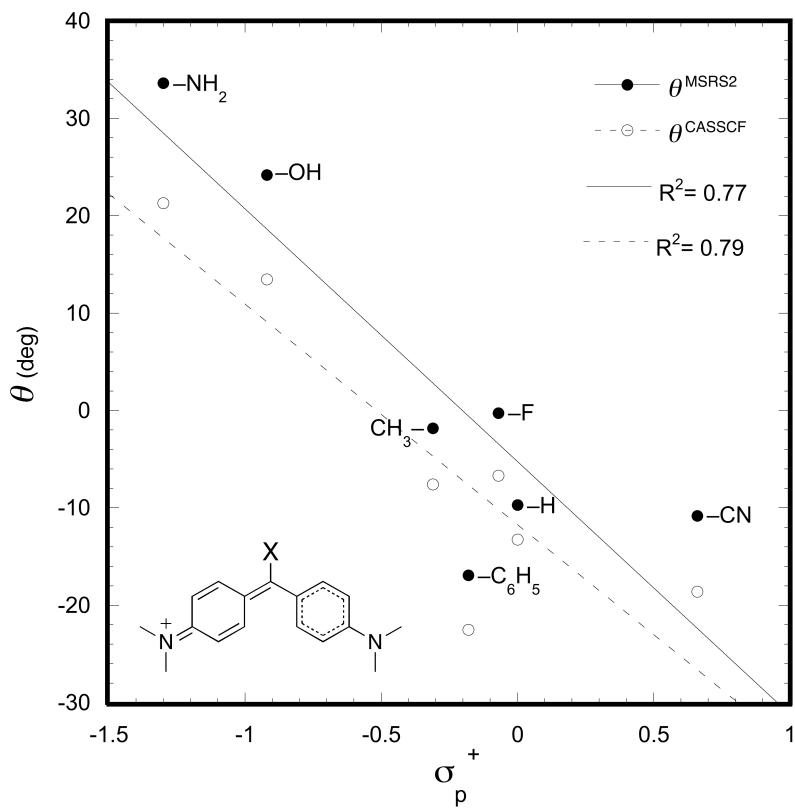

**Figure 7.** Comparison of the mixing angle $\theta$ (c.f Equation 2) with the Brown-Okamoto substituent parameter $\sigma_p^+$.

If the parameter fits for $\varepsilon$ and $\tau$ obtained in Figures 3 and 4 are inserted into the model Hamiltonian in Equation (1), and the adiabatic excitation energies calculated from this, one obtains a prediction that can be tested. Specifically, the $S_0$-$S_1$ excitation energy should begin to rise again with increasingly negative $\theta$. To the extent that the angle $\theta$ is determined by the electrodonation/withdrawal properties of the substituent, this suggests that a hypsochromic shift should be observed for substituents that are significantly more electron-withdrawing than –CN. This would amount to a violation of the Dewar-Knott rule for the absorption of these dyes, which says that electron-withdrawing substituents on the central methine lead to a lower first excitation energy[35, 52, 53]

Figure 5 shows that the adiabatic state $|S_2\rangle$ is dominated by $|B\rangle$, particularly for dyes with $\theta < 0$. Our analysis of the structure of state $|B\rangle$ indicates that it contains a significant singlet biradical interaction between the heterocyclic rings on either end of the dye. Since the rings are separated in space, the associated exchange interaction should be small, leading to a small splitting between single and triplet states with the same diradical and enhanced intersystem crossing. Hence, another strategy for experimental testing of our framework would be to investigate the relative acceleration of intersystem crossing in $S_2$ relative to $S_1$.

We have shown that the intermediate state $|B\rangle$ interpolates between two limiting states $|B-\rangle$ and $|B+\rangle$, which we suggest are approximately given as the minus and plus superpositions of the configurations $|lb^2r\rangle$ and $|l^2r^2\rangle$. The data in Figures 3 and 4 suggests that the energy difference between these limiting cases changes dramatically upon the application of the perturbation theory treatment. This is consistent with these states being coupled through a two-electron excitation. When the low-lying states are represented in canonical orbital configurations, there is a low-energy double electron excitation that mixes into both the $|S_0\rangle$ and $|S_2\rangle$ states. We have previously argued that the presence of this low-lying double excitation can be anticipated by considering the Hückel-theoretic model of the corresponding odd-alternant hydrocarbon reference.[37] Our results here are consistent with a lowering of the energy of the double excitation when the electronic correlation model is improved, such as also occurs for the dark $2^1A_g$ state of polyenes.[54]

We emphasize that it is not at all obvious to us why the angle that characterizes the remixing of $|B\rangle$ should be the same as the angle that characterizes the mixing of $|B\rangle$ with $|L \pm R\rangle$. This appears to be a property of the self-consistent field solution that we have obtained for the dyes in this set, and we conjecture that it may be characteristic of other dyes in this class. The remixing of $|B\rangle$ suggested by Figures 3 and 4 suggests that at least a four-state model would be needed in order to achieve matrix elements that were approximately constant over the set (as is often useful in diabatic state representations[25] of the electronic structure of a single molecule over multiple geometries). We have attempted to obtain analogous solutions to the evenly-weighted four-state problem (i.e. SA4-CAS(4,3)), but have found for each case that the solution either does not converge or converges to a symmetry-broken solution. Non-evenly weighted four-state solutions can be found which have the same orbital structure as in Figure 1. However, the interpretation of these solutions using methods we employ here is not straightforward, because the diagonal form of the state-averaged ensemble density matrix would not be preserved by a unitary diabatization on the target space[41]. Some additional conceptual apparatus needs to be developed before these solutions can be meaningfully discussed. We leave this for later work.

We note that the isolobal analogy established by the SA3-CAS(4,3) solutions (c.f. Figure 1) also establishes an analogy between these dyes and the chromophore of the green fluorescent protein.[37, 38, 43] A similar, even broader, analogy can also be found within the active space of the analogous two-state SA2-CAS(4,3) solutions.[35, 36, 55] Both these dyes and GFP chromophores display a common binding-dependent fluorescence enhancement associated with the suppression of a common double-bond photoisomerization chemistry in the excited state.[4-7, 56-58] Our ongoing work shows that this analogy can also be applied to electronic structure along the photoisomerization coordinates of some of the dyes studied here. We will report on these analogies in future work. It does seem to us now that there is a general physics underlying the excited state behavior in all of these systems, and an opportunity for the development and application of quite general model Hamiltonians for fluorogenic monomethine dyes.

In our previous study of the photoisomerization behavior of the anionic GFP chromophore, we showed that the photoisomerization along two possible coordinates was correlated with a remixing of the state $|B\rangle$ with one or the other of the states $|L\rangle$, $|R\rangle$.[43] It is reasonable to ask if the change we report here for the structure of the state $|B\rangle$ will affect the nature of the photoisomerization coordinate. Preliminary results suggest the answer is affirmative. We will take this up again in future work.

## 5. Conclusion

We have expressed a four-electron, three-orbital state-averaged complete active space self-consistent field and multireference perturbation theory model for cationic diarylmethanes[38] in a diabatic form, This structure of the diabatic states are analogous to the valence-bond structures invoked in resonance-theoretic models of the low-energy spectra of these dyes. The model predicts that the "intermediate" state mediating superexchange between the canonical resonating structures can be negative, even though the charge in the resonating states is positive. The sign of the charge carrier in the intermediate state is not the same for all dyes in the family. The evolution of the sign of the charge carrier can be described using a one-parameter family of model Hamiltonians. Future work will investigate the consequences of the change in the charge carrier for photochemical processes in the low-energy excited states.

**Acknowledgement** This work has been supported by Australian Research Council Discovery Projects DP110101580 and DP08877875. Computational resources on the NCI National Facility was provided by the National Computational Merit Allocation Scheme (projects m03 and n62). We thank A. Painelli and B. Salter-Duke for helpful discussions.

**Supporting Information Available** Cartesian coordinates (Å) of molecular geometries, SA-CASSCF and MRS2 state energies (au), MSRS2 mixing matrices and SA-CASSCF natural orbitals and occupation numbers are available.

**References**

[1]     R. Rose, *J. Chem. Educ.* **3** (1926) 973.
[2]     J. Griffiths, *Colour and constitution of organic molecules* (Academic Press, London, 1976). pp. 240-270.
[3]     M. Kondo, I. A. Heisler, and S. R. Meech, *Faraday Discuss.* **145** (2010) 185.
[4]     C. Kitts, T. s. Beke-Somfai, and B. Nordén, *Biochemistry* **50** (2011) 3451.
[5]     C. Szent-Gyorgyi, B. Schmidt, Y. Creeger, G. Fisher, K. Zakel, S. Adler, J. Fitzpatrick, C. Woolford, Q. Yan, K. Vasilev, P. Berget, M. Bruchez, J. Jarvik, and A. Waggoner, *Nature Biotechnol.* **26** (2008) 235.
[6]     J. R. Babendure, S. R. Adams, and R. Y. Tsien, *J. Am. Chem. Soc.* **125** (2003) 14716.
[7]     R. H. Conrad, J. R. Heitz, and L. Brand, *Biochemistry* **9** (1970) 1540.
[8]     K. Eisenthal, *Chem. Rev.* **106** (2006) 1462.
[9]     P. Sen, S. Yamaguchi, and T. Tahara, *Faraday Discuss.* **145** (2010) 9.
[10]    S. Rafiq, R. Yadav, and P. Sen, *J. Phys. Chem. B* **114** (2010) 13988.
[11]    P. Changenet, H. Zhang, M. Van der Meer, M. Glasbeek, P. Plaza, and M. Martin, *J. Phys. Chem. A* **102** (1998) 6716.
[12]    D. Kolpashchikov, *J. Am. Chem. Soc.* **127** (2005) 12442.
[13]    M. Du, Y. Jia, and G. Fleming, *Springer Series in Chemical Physics* **50** (1994) 515.
[14]    M. Vogel, and W. Rettig, *Ber. Bunsen. Phys. Chem.* **89** (1985) 962.

[15] L. Brey, G. Schuster, and H. Drickamer, *J Chem Phys* **67** (1977) 2648.
[16] M. Van der Meer, H. Zhang, and M. Glasbeek, *J Chem Phys* **112** (2000) 2878.
[17] L. Pauling, *Proceedings of the National Academy of Sciences* **25** (1939) 577.
[18] W. Moffitt, *Proc. Phys. Soc. A* **63** (1950) 700.
[19] L. Brooker, *Rev. Mod. Phys.* **14** (1942) 275.
[20] J. Platt, *J Chem Phys* **25** (1956) 80.
[21] D. Lu, G. Chen, J. Perry, and W. Goddard III, *J. Am. Chem. Soc.* **116** (1994) 10679.
[22] W. Thompson, M. Blanchard-Desce, and J. Hynes, *J. Phys. Chem. A* **102** (1998) 7712.
[23] F. Terenziani, O. Przhonska, S. Webster, L. Padilha, Y. Slominsky, I. Davydenko, A. Gerasov, Y. Kovtun, M. Shandura, A. Kachkovski, and A. Painelli, *J. Phys. Chem. Lett.* **1** (2010) 1800.
[24] L. Grisanti, G. D'Avino, A. Painelli, J. Guasch, I. Ratera, and J. Veciana, *J. Phys. Chem. B* **113** (2009) 4718.
[25] T. Van Voorhis, T. Kowalczyk, B. Kaduk, L.-P. Wang, C.-L. Cheng, and Q. Wu, *Annu. Rev. Phys. Chem.* **61** (2010) 149.
[26] S. Shaik, *New J. Chem.* **31** (2007) 1981.
[27] S. Shaik, and P. C. Hiberty, *WIREs Comput. Mol. Sci.* **1** (2011) 18.
[28] W. Simpson, *J. Am. Chem. Soc.* **75** (1953) 597.
[29] A. E. Azhary, G. Rauhut, P. Pulay, and H.-J. Werner, *J Chem Phys* **108** (1998) 5185.
[30] T. H. Dunning, *J Chem Phys* **90** (1989) 1007.
[31] I. Baraldi, A. Carnevali, and F. Momicchioli, *Chem. Phys.* **160** (1992) 85.
[32] C. Aaron, and C. Barker, *J. Chem. Soc.* (1963) 2655.
[33] H.-J. Werner, and W. Meyer, *J Chem Phys* **74** (1981) 5794.
[34] P. Pulay, and T. P. Hamilton, *J Chem Phys* **88** (1988) 4926.
[35] S. Olsen, *Chem Phys Lett* ( http://dx.doi.org/10.1016/j.bbr.2011.03.031)
[36] S. Olsen, and R. H. McKenzie, *J Chem Phys* **134** (2011) 114520.
[37] S. Olsen, and R. H. McKenzie, *Chem Phys Lett* **492** (2010) 150.
[38] S. Olsen, *J. Phys. Chem. A* **116** (2012) 1486.
[39] T. Pacher, L. Cederbaum, and H. Köppel, *J Chem Phys* **89** (1988) 7367.
[40] J. Stålring, A. Bernhardsson, and R. Lindh, *Mol. Phys.* **99** (2001) 103.
[41] L. Hughston, R. Jozsa, and W. Wootters, *Phys. Lett. A* **183** (1993) 14.
[42] R. Cave, and M. Newton, *J Chem Phys* **106** (1997) 9213.
[43] S. Olsen, and R. H. McKenzie, *J Chem Phys* **130** (2009) 184302.
[44] J. Campo, A. Painelli, F. Terenziani, T. Van Regemorter, D. Beljonne, E. Goovaerts, and W. Wenseleers, *J. Am. Chem. Soc.* **132** (2010) 16467.
[45] R. Hoffmann, *Angew. Chem. Int. Ed.* **21** (1982) 711.
[46] J. E. Subotnik, S. Yeganeh, R. J. Cave, and M. A. Ratner, *J Chem Phys* **129** (2008) 244101.
[47] J. Finley, P. Å. Malmqvist, B. O. Roos, and L. Serrano-Andrés, *Chem Phys Lett* **288** (1998) 299.
[48] C. Angeli, R. Cimiraglia, and J. Malrieu, *J. Chem. Educ.* **85** (2008) 150.
[49] B. Roos, K. Andersson, M. Fülscher, L. Serrano-Andrés, K. Pierloot, M. Merchán, and V. Molina, *J. Mol. Struct.: THEOCHEM* **388** (1996) 257.
[50] J. Koutecký, J. Paldus, and J. Čížek, *J Chem Phys* **83** (1985) 1722.
[51] H. Brown, and Y. Okamoto, *J. Am. Chem. Soc.* **80** (1958) 4979.
[52] M. Dewar, *J. Chem. Soc.* (1950) 2329.
[53] E. Knott, *J. Chem. Soc.* (1951) 1024.
[54] P. Tavan, and K. Schulten, *J Chem Phys* **70** (1979) 5407.
[55] S. Olsen, *J. Chem. Theory Comput.* **6** (2010) 1089.
[56] L. M. Tolbert, A. Baldridge, J. Kowalik, and K. M. Solntsev, *Acc. Chem. Res.* **45** (2012) 171.
[57] S. R. Meech, *Chem. Soc. Rev.* **38** (2009) 2922.
[58] S. Remington, *Curr. Opin. Struc. Biol.* **16** (2006) 714.